\documentclass{aastex62}
\usepackage{natbib}
\usepackage{gensymb}
\usepackage{float}
\usepackage{graphicx}
\usepackage{color}
\usepackage{amsmath}
\usepackage{ulem}
\usepackage{breqn}
\usepackage{textgreek,stackengine}
\newcommand\textdot[1]{\stackon[1pt]{\csname text#1\endcsname}{.}}

\begin{document}

\title{Gamma-ray Bursts as Distance Indicators by a Statistical Learning Approach}


\correspondingauthor{Maria Giovanna Dainotti}
\email{maria.dainotti@nao.ac.jp}

\author{Maria Giovanna Dainotti}
\affiliation{National Astronomical Observatory of Japan, Mitaka, Tokyo 181-8588, Japan;}
\affiliation{The Graduate University for Advanced Studies, SOKENDAI, Kanagawa 240-0193, Japan;}
\affiliation{Space Science Institute, Boulder 80301, CO, USA;}

\author{Aditya Narendra}
\affiliation{Doctoral School of Exact and Natural Sciences, Jagiellonian University, Krakow, Poland;}
\affiliation{Astronomical Observatory of Jagiellonian University, Krakow, Poland;}

\author{Agnieszka Pollo}
\affiliation{Astronomical Observatory of Jagiellonian University, Krakow, Poland;}
\affiliation{National Center for Nuclear Physics (NCB), Warsaw;}

\author{Vah\'e Petrosian}
\affiliation{Department of Physics  Stanford University, 382 Via Pueblo Mall, Stanford, CA 94305-4060;}
\affiliation{Kavli Institute for Particle Astrophysics and Cosmology, Stanford University;}
\affiliation{Department of Applied Physics, Stanford University;}

\author{Malgorzata Bogdan}
\affiliation{ Department of Mathematics, University of Wroclaw, 50-384, Poland;}
\affiliation{Department of Statistics, Lund University, SE-221 00 Lund, Sweden;}

\author{Kazunari Iwasaki}
\affiliation{National Astronomical Observatory of Japan, Mitaka, Tokyo 181-8588, Japan;}
\affiliation{The Graduate University for Advanced Studies, SOKENDAI, Kanagawa 240-0193, Japan;}
\affiliation{ Center for Computational Astrophysics, National Astronomical Observatory of Japan, 2 Chome-21-1 Osawa, Mitaka, Tokyo 181-8588, Japan;}

\author{Jason Xavier Prochaska}
\affiliation{University of California, Santa Cruz, 1156 High Street, Santa Cruz, CA 95064;}

\author{Enrico Rinaldi}
\affiliation{Interdisciplinary Theoretical and Mathematical Sciences (iTHEMS) Program, RIKEN, Wakoshi, Saitama 351-0198, Japan;}

\author{David Zhou}
\affiliation{Arizona State University, 1151 S Forest Ave Tempe, AZ 85281;}

\begin{abstract}
Gamma-ray bursts (GRBs) can be probes of the early universe, but currently, only 26\% of GRBs observed by the Neil Gehrels Swift Observatory GRBs have known redshifts ($z$) due to observational limitations. 
To address this, we estimated the GRB redshift (distance) via a supervised {statistical} learning model that uses optical afterglow observed by Swift and ground-based telescopes. 
The inferred redshifts are strongly correlated (a Pearson coefficient of 0.93) with the observed redshifts, thus proving the reliability of this method.
The inferred and observed redshifts allow us to estimate the number of GRBs occurring at a given redshift (GRB rate) to be {8.47-9}$yr^{-1} Gpc^{-1}$ for $1.9<z<2.3$.
Since GRBs come from the collapse of massive stars, we compared this rate with the star formation rate highlighting a discrepancy of a factor of 3 at $z<1$.
\end{abstract}

\section{Introduction}\label{intro}

GRBs, because of their high luminosity are detected up to redshift {$z=9.4$}, and thus can be vital cosmological probes of early Universe. 
Since they cover a wide redshift range, {$\sim 0.09<z<9.4$,} they can also map the evolution of the universe; the cosmic star formation rate (SFR),  
the formation of first generation (Population III) stars, and the birth of black holes.
Tracking this history requires a large sample of GRBs with known redshifts and observational selection criteria.
With such samples we can determine the GRB luminosity function (LF) and their 
formation rate (FR) evolution per co-moving volume. 
Additionally, there have been attempts to use GRBs as ``standard candles" \citep{amati2002,Ghirlanda2004ApJ...616..331G,yonetoku2004,Dainotti2008} using correlations, between a distance-dependent and a distance-independent variable, to extend the Hubble diagram to higher redshifts. 

A persistent problem  for study of either cosmological evolution or determination of  correlations has been the incompleteness of the GRB samples with known redshifts due to various observational selection effects, which truncate data and introduce bias for both tasks, especially the evolution with redshift of the correlations.
Additionally, the broad distributions of most intrinsic characteristics of GRBs exacerbate the problem.

These broad distributions hinder the search for an effective strategy to derive redshifts reliably from the observational predictors. 
It is essential to determine how these distributions vary with the redshift, namely, the functional form or their cosmological evolution \citep{Dainotti2013a,petrosian2015,dainotti2015luminosity}, before attacking any of the above problems.
To determine these functional forms, we need large samples with known redshifts and well-defined observational selection criteria. This allows for a precise determination of cosmological evolutions.

Determining cosmological evolution is challenging, but several parametric and non-parametric approaches, have been proposed to overcome this issue.
In parametric approaches, a set of assumed \textit{parametric} functional forms are used to fit the data to determine the ``best-fit values'' of the parameters. 
This involves assumed forms for several functions: the LF, density rate evolutions, photon spectrum, light curve, etc., each with three or more parameters. 
Because of the large number of parameters, there is no certainty about the uniqueness of the results. 
Moreover, these methods often require binning and, hence, large samples. 

There have been many such studies with these methods of both long-duration GRBs (LGRBs, see, e.g., \cite{porciani2001,jakobsson2006mean,Dong2022MNRAS.513.1078D,Ghirlanda2022ApJ...932...10G} and references therein) 
and short duration GRBs (SGRBs, see, e.g., \cite{Nakar2006ApJ...650..281N,Virgili2009MNRAS.392...91V,metzger2012most,Petrillo2013ApJ...767..140P,Paul2018MNRAS.477.4275P,Dainotti2021ApJ...914L..40D}).
LGRBs have $T_{90} >2$, where $T_{90}$ \citep{mazets1981catalog,kouveliotou1993identification} is the time in which a burst emits $90\%$ of its total background-subtracted counts between the 5 and 95\%, while SGRBs have $T_{90} < 2$s.


Interestingly, \cite{petrosian2015,Yu2015ApJS..218...13Y,pescalli2016A&A...587A..40P} and \cite{tsvetkova2017konus} discovered a difference between the evolution of the LGRB and the global SFR at low redshifts ($0<z<1$). 
However, \cite{pescalli2016A&A...587A..40P} found smaller differences between GRB and star FR. 
We focus on this ongoing debate and determine the true GRB rate.
Our goal with the { statistical learning methodology} is to increase the number of LGRBs with inferred redshift considerably 
(we will almost double the sample of {\it Swift GRBs} with $z$ if we consider both X-rays and optical data) 
so that we can solve the above debate.

Almost doubling the sample will allow us to answer more accurately the question of the true shape of the LF and to what extent the GRB rate follows the star FR (SFR).

The problem of the incompleteness of the sample is a critical issue both for LGRBs and SGRBs
\citep{Dainotti2021ApJ...914L..40D,Paul2018MNRAS.477.4275P,Wanderman2015MNRAS.448.3026W}.
Accurately determining the redshift for SGRBs is crucial with the dawn of gravitational wave astronomy.
Therefore, a method that can reliably estimate redshifts for more GRBs both at low and high-$z$ is paramount.
At high-$z$, we can trace the SFR of Population III.
Since only a small fraction of GRBs observed by Swift ($\sim 26\%$) have redshift, we need methods to infer the redshifts. 

As mentioned above, using correlations 
has allowed us to obtain {pseudo-redshifts} for larger samples for cosmological studies 
\citep{Lloyd_Ronning_2022,atteia2003,yonetoku2004,Kocevski2006ApJ...642..371K,Tsutsui2011PASJ...63..741T,Dainotti2011ApJ...730..135D,zhang2018}, but the inferred redshifts are not yet accurate.
{
Recently, \cite{Razaque2024MNRAS.529.2676A} trained a Deep Neural Networks to predict the redshift of Fermi-GBM and Konus Wind GRBs using spectral fit parameters. However, they do not perform cross-validation and do not eliminate duplicated GRBs from the two data sets.
Regardless, although not reliable, their results do highlight the strength of such an approach.}
We propose a very effective method using { statistical learning}  tools for determining redshifts.

In \S\ref{sample}, we detail our GRB sample used for the redshift inference procedure via methods, described in \S\ref{Methods}. 
In \S\ref{results}, we describe how the chosen procedures reproduce the observed spectroscopic redshifts. 
In \S \ref{evol}, we compare the results from applying { statistical learning} methods to both predicted redshift ($z_{pred}$) and observed redshift ($z_{obs}$) to obtain the LF and density rate evolution. 
We also compare our density rate evolution with the established results in the literature.
A summary and discussion of the impact on future work are given in \S \ref{summary}.

\begin{figure}
    \includegraphics[width=\textwidth]{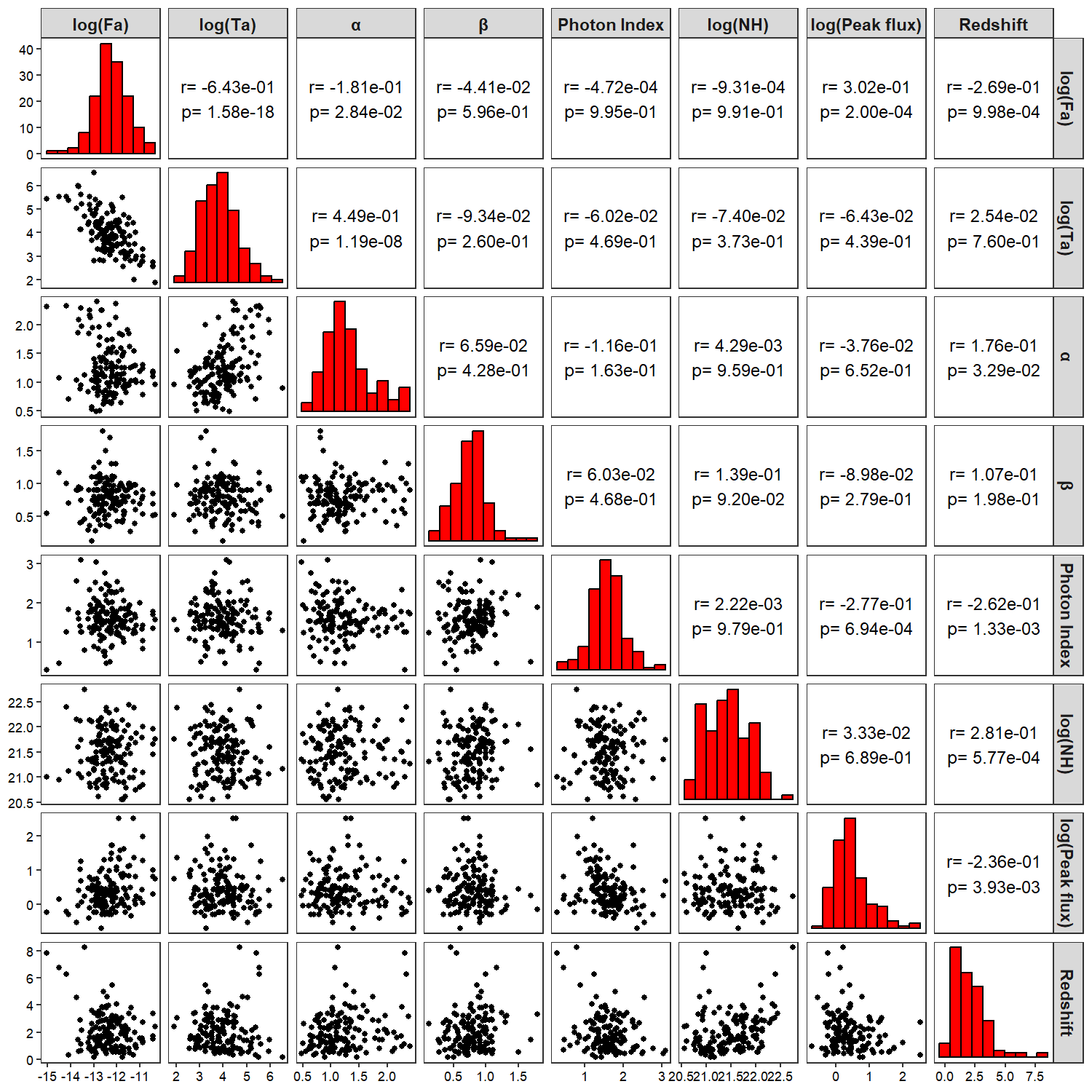}
    \caption{The scatter matrix plot of the predictors at play, some showing strong correlations, while others show weak or no correlation. 
    Each variable (the prompt $Fluence$, $T_{90}$, the prompt $Peak flux$, the plateau optical flux ($\log(F_{a})$) its duration ($\log(T_{a})$) the hydrogen column density ($\log(NH)$), the prompt $Photon$ $Index$, the temporal and spectral indices, $\alpha$, and $\beta$, of the optical afterglow and plateau emissions, respectively, 
    is plotted against every other. 
    The stars next to the correlation value denote the significance level. One, two, or three stars denote the corresponding variables' correlation is significant at 10\%, 5\% and 1\% levels, respectively.
    }
    \label{scatter}
\end{figure}

\section{The sample}\label{sample}
Our analysis adopts several observed characteristics of GRBs. 
GRBs have, historically, been divided into the classes of SGRBs and LGRBs depending on their $T_{90}$. 
SGRBs are produced by the merger of compact objects \citep{duncan1992}, and LGRBs result from the collapse of massive stars \citep{woosley1993ApJ...405..273W}.

In about 42\% of cases, the afterglow presents the plateau emission, a flat part of the GRB lightcurves, whose parameters we include as {\it predictors} of their redshifts. 

The data sets are composed of 179 GRBs with optical plateaus from the Swift Ultraviolet/Optical Telescope (UVOT) and from 455 ground-based telescopes/detectors, e.g., the Subaru Telescope, Gamma-ray Burst Optical/Near-IR Detector (GROND), Reionization and Transients InfraRed camera/telescope (RATIR), the MITSuME \citep{Kotani2005NCimC..28..755K}, observing from 1997 May to 2021 May.
Fig. \ref{scatter} shows a scatter matrix plot of all the variables (including the redshift). 
This matrix includes the logarithms of the redshift, prompt $Fluence$, $T_{90}$, the prompt $Peak flux$, the peak flux computed in 2 second, the time at the end of the plateau emission $T_{a}$, its correspondent flux, $F_{a}$, and the column density, $NH$, while the following variables in linear scale: the prompt $Photon$ $Index$, and the spectral index (assuming a simple power-law spectrum), $\beta$, of the plateau emission, and the temporal decay index after the plateau, $\alpha$. 
As it is clear from the Fig. \ref{scatter}, some of these scatter diagrams show clear linear phenomenological relations such as the bi-dimensional \cite{Dainotti2008} relation in optical \citep{Dainotti2020b} between the plateau luminosity and rest frame duration, $L_a$ and $T^{*}_a$, and its extension in three dimensions (3D) by adding the prompt peak luminosity, $L_{peak}$ \citep{Dainotti2016,dainotti2017c,dainotti2020a}, and the various spectral hardness-luminosity relations \citep{Collazzi2008ApJ...688..456C}. The bi-dimensional $L_a$ and $T^{*}_a$ relation stems from the $\log F_a- \log T_a$ relation between the flux at the end of the plateau emission, $F_a$, and its duration, $T_a$ in the observer frame, as the predictors shown in Fig.1, has a Pearson correlation of $-0.643$. The 3D correlation can be seen through the observer frame relation between the $\log Peak flux$ vs $\log F_a$. Here, the Pearson correlation is small since in the 3D relation \citep{dainotti2022} the peak flux is given in energy flux (erg/cm$^{2}$s) and not as a photon flux (photons/cm$^{2}$s) as in this work. For these relations, we acknowledge that from previous works \citep{Dainotti2013a,dainotti2015luminosity,dainotti2017b,Dainotti2020b,levine2022,dainotti2022} we have shown that these relations are intrinsic and not induced by detector threshold truncations.
For recent reviews on correlations see \cite{2017NewAR..77...23D,2018AdAst2018E...1D,dainotti2018PASP..130e1001D}.
In addition to these relations, we also note a strong relation between prompt $Fluence$ and $Peak flux$. 

\subsection{The sample data cut and the MICE algorithm}

\begin{figure}
   \centering
    \includegraphics[width=\textwidth]{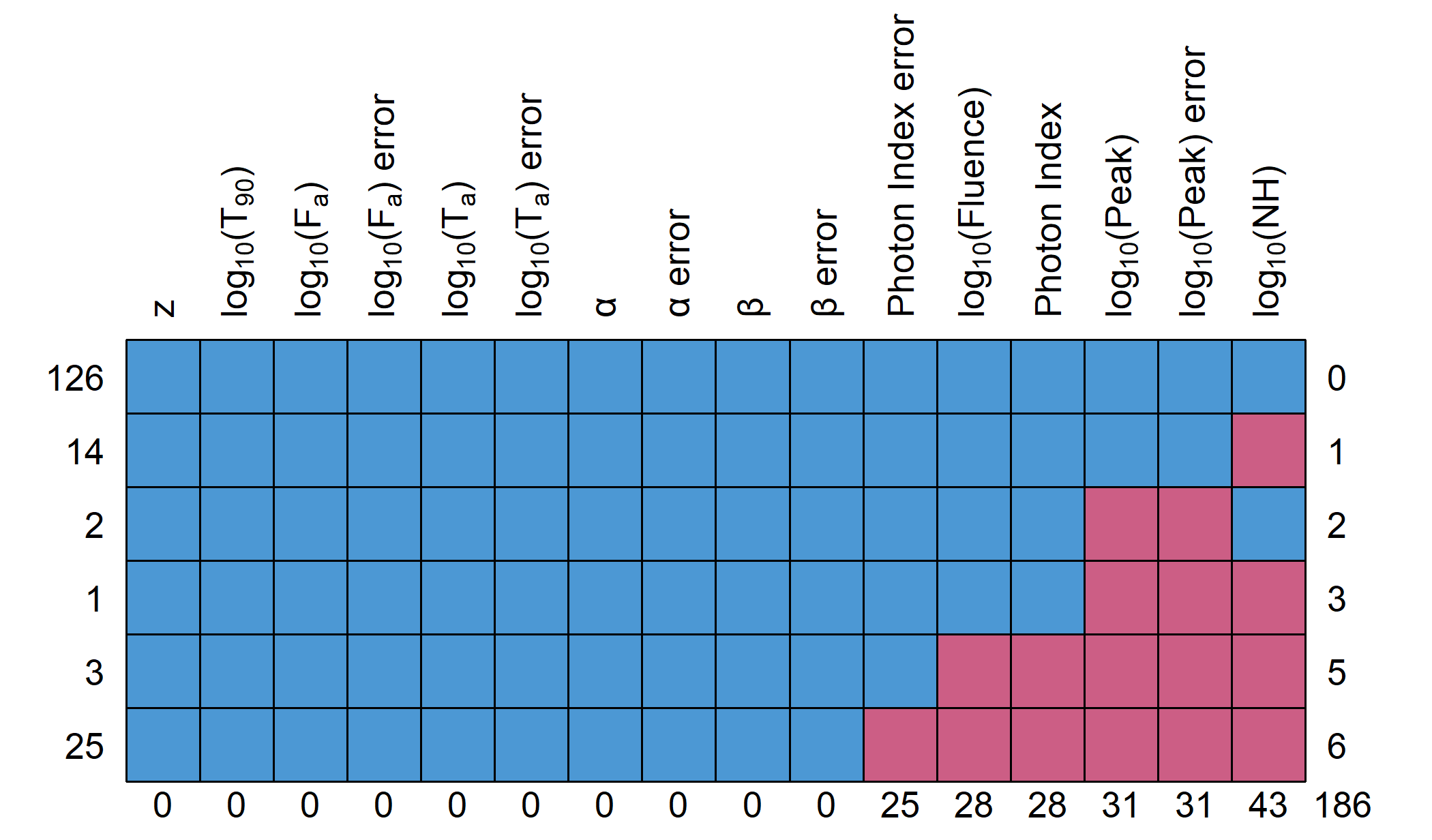}
    \caption{The missing data in our sample. The pink boxes show the missing GRBs in relation to a given variable presented in the upper x-axis, while the blue boxes indicate GRBs with no missing data for a given GRB variable presented in the top axis. The bottom axis shows the number of missing GRBs per variable. The left axis represents the number of observations that have missing data for a specific set of predictors. The right axis represents the number of { predictors} that are missing for that row. For example, starting from the first row, there are 126 GRBs with no missing data, 14 GRBs with missing data in $\log(Peak flux)$ data, 2 GRBs with missing data in $\log(NH)$, and $\log(Peak flux)$ error, and so on. 
    The corner number on the right of the X-axis (186) shows the total number of missing predictors in the data set, which is the sum of the numbers shown along the axis}.
    \label{missing_mice}
\end{figure}

From our 179 GRBs, we removed the SGRBs and Short with the extended emission GRBs, arriving to 171 GRBs containing 45 with 4 missing data in $NH$, $\log(Peak flux)$, $\log(Fluence)$, and the $Photon$ $Index$ (see Fig. \ref{missing_mice}).
We have 10 GRBs with a non-physical value of $\log(NH) < 20$ and 2 GRBs with $Peakflux = 0$. We also consider these non-physical values as missing data.

To overcome this issue and retain the largest sample for training, we have applied an imputation method, the Multivariate Imputation by Chained Equations (MICE) \citep{van2011mice}.
Imputing these values with MICE helps enhance our training set by 26\%.

MICE can impute missing values for multiple variables using the data set with complete variables.
However, the missing data has to be assumed to be missing at random (MAR) \citep{rubin1976inference,schafer2002missing}. 
Here, we use the method of predictive mean matching (PMM)\citep{little2019statistical}, named `midastouch'.
In PMM, missing {   predictors } are initialized with their mean values, and then their estimates are obtained by training a model on the rest of the complete data. 
To reduce the imputation's randomness, we will impute the missing variables 20 times for each MICE iteration instead of the usual practice of 10 times \citep{van2011mice}.
We tested the imputed and original distributions with the Kolmogorov-Smirnov test, which yielded a significance level of 0.5.
Previously, \cite{gibson2022} applied MICE to impute missing values of Active Galactic Nuclei.
There, we demonstrated that this imputation method does not add any additional uncertainty to the variable's original distribution and helps strengthen the final training set used for the { statistical learning} application by enlarging it.

After the imputation, we split the data into training and test sets in a 90:10 split.
Next, we make three additional cuts where we remove outliers that have $\log(F_a)>-10$, $\alpha>2.5$, and $\log(NH)<20.5$.
Finally, we arrive at 150 GRBs, with which we perform the M-estimator.



\section{Results}\label{results}
\subsection{The Statistical learning results}
\begin{figure}[H]
\includegraphics[width = 0.49\textwidth]{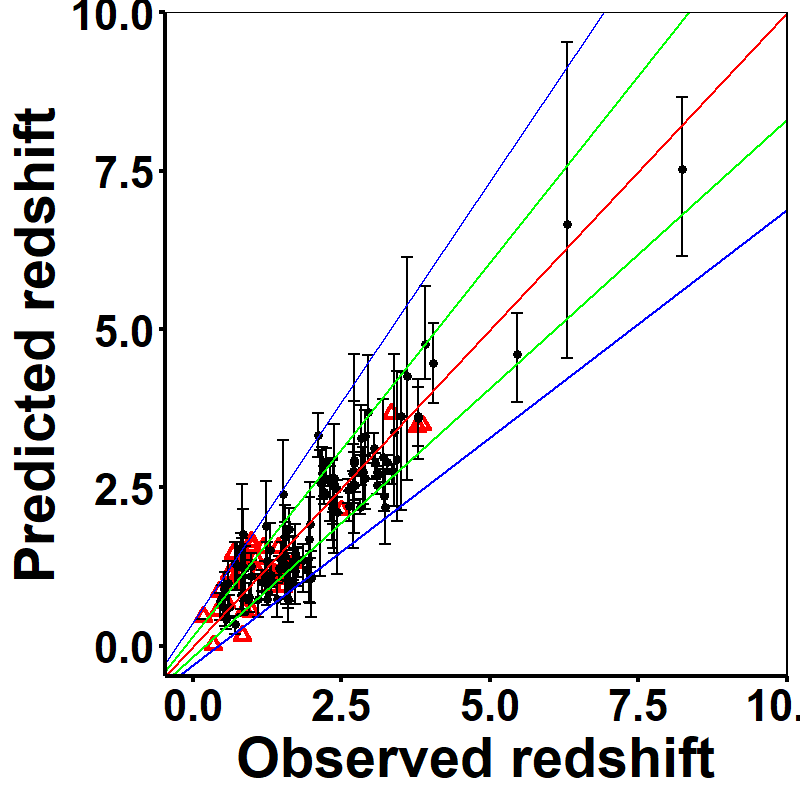}
\includegraphics[width = 0.49\textwidth]{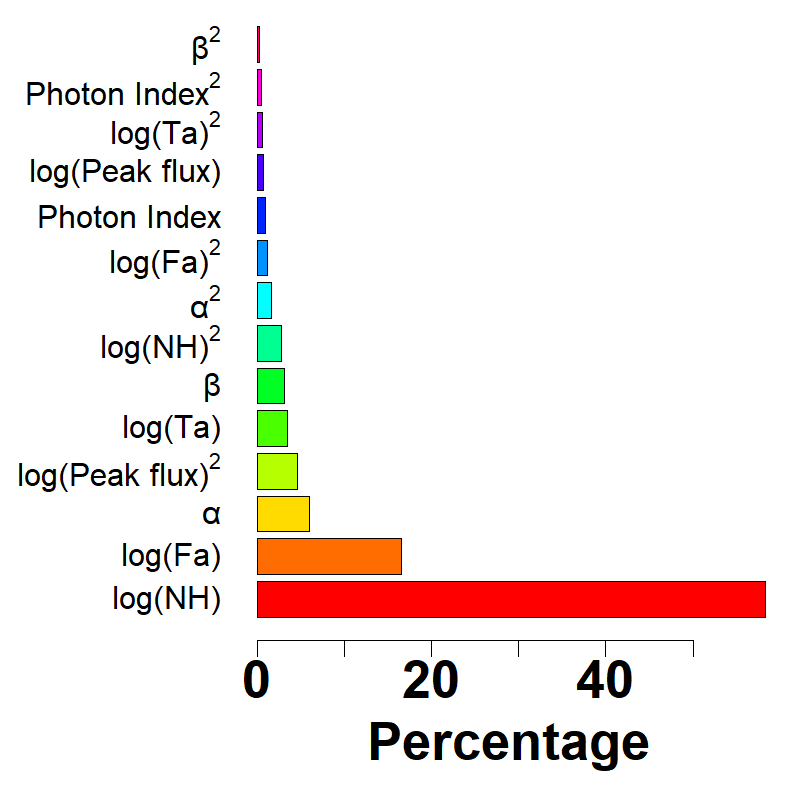}
\caption{Left panel: The $z_{obs}$ vs $z_{pred}$ scatter plot for 103 GRBs. The red triangles indicate the GRBs with $\delta z_{pred}/z_{pred}$$>$1.
Right panel: The relative influence of each variable on the predicted results.}
\label{influence}
\end{figure}

We present the results derived with the ensemble method, the SuperLearner, which has the advantage of combining several { statistical learning} methods into a single model and leveraging the predictive power of each model.
This is the first time SuperLearner has been applied to GRBs.
SuperLearner computes coefficients for each learner based on their performance within the ensemble. 
Each model and the Superlearner are discussed in \S \ref{ML}.
In the left panel of Fig. \ref{influence}, we show the observed redshift ($z_{obs}$) versus the predicted redshift ($z_{pred}$) via a 10-fold cross-validation technique averaged over 100 runs (see \S \ref{SuperLearner}). 
The average mean squared error, MSE, is defined as the $<(z_{pred} - z_{obs})^2>$ where the symbol $<>$ indicates the average between $z_{pred}$ vs $z_{obs}$. 
The obtained MSE is $0.21$, with a high Pearson correlation ($r$) of $0.93$ and a very small average bias, defined as the mean $<(z_{pred}- z_{obs})>$ of $0.02$. 

The red line shows the equality line, while the green and blue lines show the cone of $|{\Delta z}| > 1\sigma$ and 2$\sigma$ (see \S \ref{errorbarSection} for derivation), respectively.
The red triangles indicate GRBs with $\delta z_{pred}/z_{pred} \geq 1$, and thus were removed them from the LF and density rate analysis.
Here $\Delta z$ is $z_{obs}-z_{pred}$,
while $\delta z_{pred}$ is the error bar on $z_{pred}$.


In the right panel of Fig. \ref{influence} we have shown the relative influence of the variables obtained with SuperLearner (see \S \ref{SuperLearner} for details).
The relative importance of predictors is an average of the importance given by a local linear approximation of predictions. 
The variable $\log(NH)$ obtained the highest influence in predicting the redshift. 
This variable gives the NH obtained from an automated time-averaged spectral fit of the Swift XRT data \citep{Evans2009}.
The high influence of this variable can be understood
because the more distant a GRB is, the more neutral hydrogen it will encounter. 
Thus, the redshift of a GRB depends on the calculated $\log(NH)$ value.
The second, third, fourth and fifth most predictive variables are $\log(F_a)$, $\alpha$, $\log(Peak flux)$, $\log(T_a)$. 
The variables related to prompt and the plateau emission, such as the time at the end of the plateau ($log(T_a)$) and its correspondent flux $\log(F_a)$, and the prompt $Peak flux$ are among the best predictors, thus recovering the key parameters of the three-dimensional Dainotti correlation (named the GRB fundamental plane relation), the correlation between the peak prompt luminosity, the luminosity at the end of the plateau emission and the rest end time of the plateau emission \citep{Dainotti2016,dainotti2017b,dainotti2020a,Dainotti2020b,dainotti2022}.

Since $\log(T_a)$ and its flux are among the most important predictors it is relevant to discuss the theoretical explanation of the plateau emission \citep{Zhang2001} and how the Dainotti relations \citep{Rea2015ApJ...813...92R,Stratta2018} results from the magnetar model. 
This model is based on the dipole radiation emitted by the rotational energy of a newly born neutron star (NS) \citep{thompson1994MNRAS.270..480T}. 
The plateau ends when the NS reaches its critical spin-down timescale; the uncertainties on $T_a$ can be ascribed to the uncertainties on the magnetar spin period and magnetic field.
Due to its theoretical explanation, the plateau is easier to use as a standard feature than the prompt's varied properties.
This explains the great advantage of our approach, which includes afterglow plateau parameters for the first time for estimating GRB redshifts, differently from other studies (e.g., \cite{morgan2012,ukwatta2016machine}).

Compared to the results obtained by using only Random Forest by \cite{ukwatta2016machine}, {the predictive power of SuperLearner has a {$\mathbf{61\%}$ increase in the $r$} coefficient between $\mathbf{z_{pred}}$ and $\mathbf{z_{obs}}$} ($r=0.93$ vs. $r=0.57$) even though we use fewer number of predictors ($7$ vs $11$) compared to \cite{ukwatta2016machine}.


\subsection{Luminosity function and density rate evolution}\label{evol}
To check whether our results can be used for deriving important astrophysical properties such as the GRB LF and the density rate evolution, we applied the Efron \& Petrosian \citep{Efron1992} method to derive these properties from both $z_{obs}$ and $z_{pred}$.
To have more precise results, we selected 103 GRBs with $\delta z_{pred}/z_{pred} < 1$. 
These GRBs are presented in the left panel of Fig. \ref{influence}.
We have analyzed the LF and density rate evolution using the method presented by \cite{petrosian2015} and \cite{Dainotti2013,dainotti2015}. 
According to their prescription, the sample of 103 GRBs was further trimmed to 97 GRBs based on a flux limit of $1.25\times10^{-14}$ erg cm$^{-2}$ (see \S \ref{EPmethod} and Fig.\ref{luminosities}). 

\begin{figure}[h!]	
\centering
\includegraphics[width=0.49\textwidth]{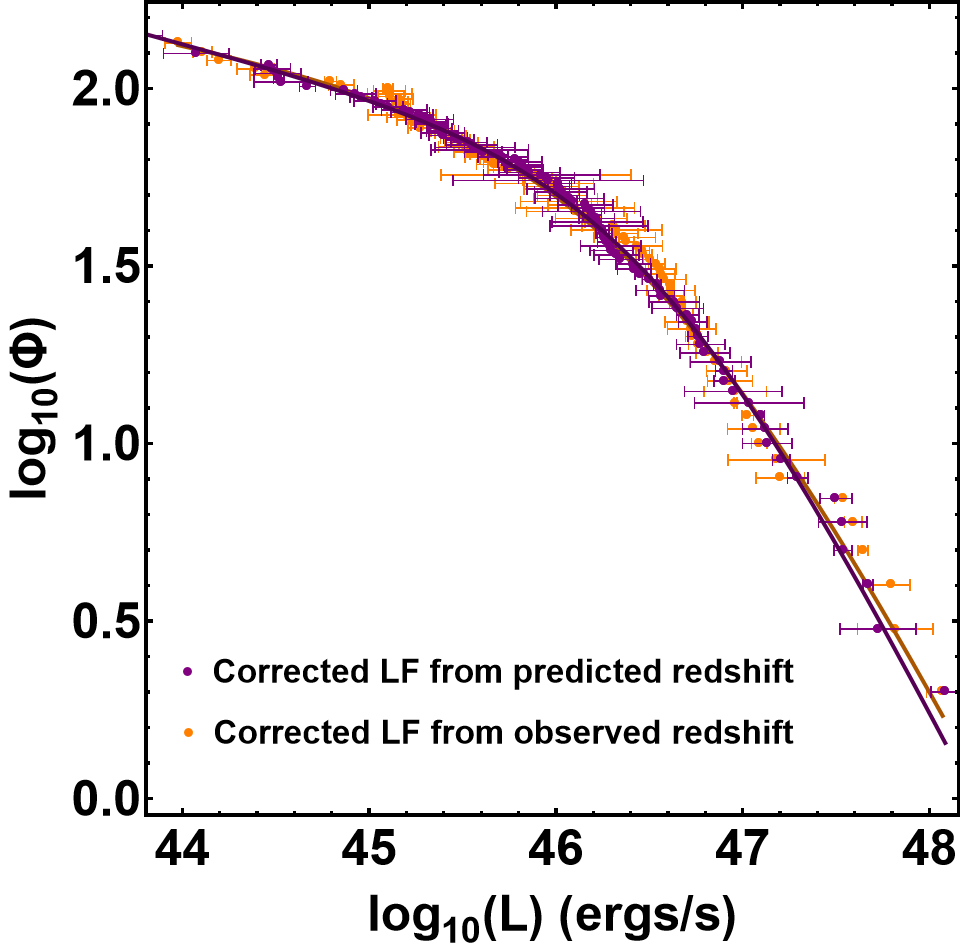}
\includegraphics[width=0.49\textwidth]{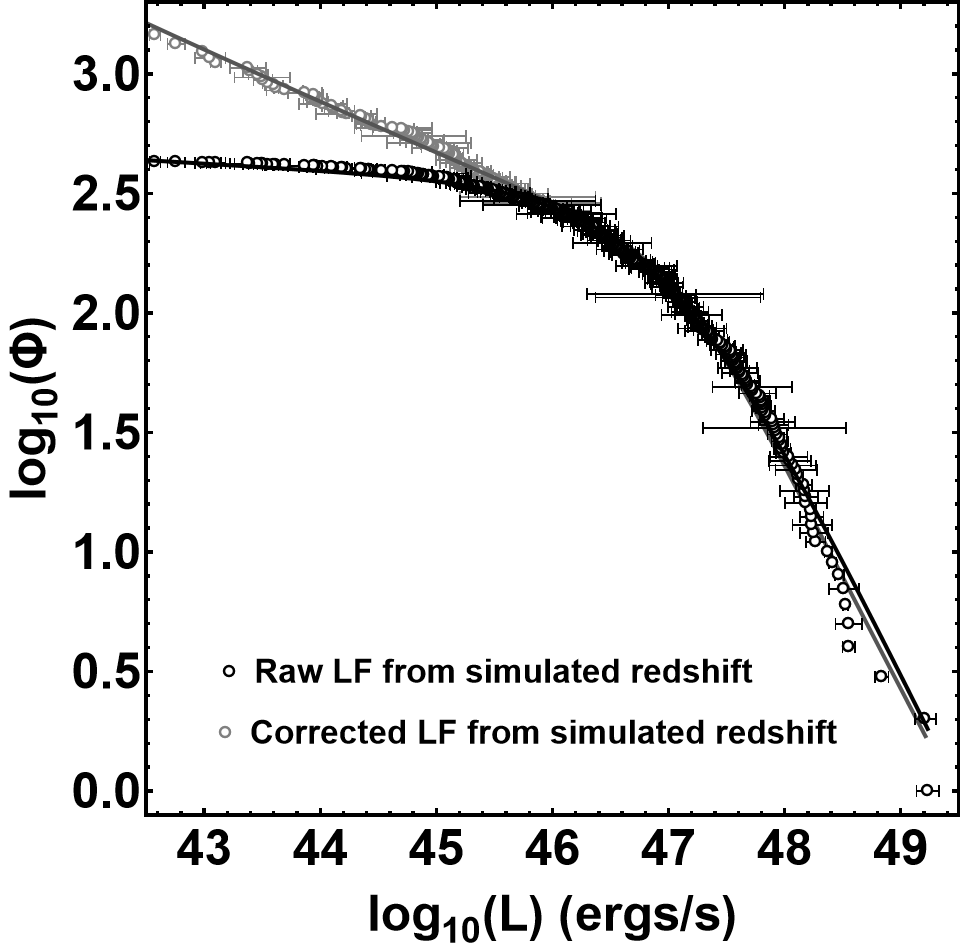}
 \caption{\small 
 Left Panel: The corrected LF. The orange and purple dots represent the corrected LF obtained from $z_{obs}$ and $z_{pred}$, respectively. 
 Right Panel: The LF combined from the $z_{obs}$ and the simulated $z_{pred}$. The black and gray points show the raw LF distribution and the EPL corrected LF, respectively. 
 The black and gray lines show the SBPL fits for the raw and corrected LF, respectively.
 }
\label{fig:lf}
\end{figure}

The left panel of Fig. \ref{fig:lf} shows the LF corrected for selection biases and redshift evolution. 
We also calculated the raw (the observed, uncorrected for redshift evolution and biases) LF.
The Smooth Broken Powerlaw (SBPL) fit is shown with orange and purple lines for the corrected LF for $z_{obs}$ and $z_{pred}$, respectively.
To guarantee that the raw and corrected LF from $z_{pred}$ and $z_{obs}$ come from the same parent population, we computed the Kolmogorov-Smirnov test. 
Both p-values of the raw and corrected LF are 0.99, showing that LF derived from $z_{obs}$ and $z_{pred}$ are from the same distribution. 

The log luminosity break for the raw SBPL for $z_{obs}$ is $46.42 \pm 0.14$, and for $z_{pred}$ is at $46.7 \pm 0.16$.
Similarly, for the corrected LF, the log luminosity break for the SBPL for $z_{obs}$ is at $46.6 \pm 0.3$, and for $z_{pred}$ is at $46.6 \pm 0.11$.
Both the luminosity breaks agree within 1$\sigma$ for the raw and corrected cases.
These results prove the reliability of using $z_{pred}$ for the analysis of the LF, demonstrating the strength of {statistical learning-based approaches}.
The right panel of Fig. \ref{fig:lf} shows the LF derived using a Markov Chain Monte Carlo (MCMC) simulation using the distributions of $z_{obs}$ and $z_{pred}$.
The black circles show the raw LF obtained with the combined sample of observed and simulated luminosities. 
The gray circles show the LF corrected for selection biases and redshift evolution.
The black and gray lines show the SBPL fits for the raw and corrected LF, respectively.
The luminosity breaks obtained are $46.9\pm0.03$ and $47.25\pm0.03$ for the raw and corrected LF, respectively. We here compute the plateau luminosity so that we can have an independent measure of the LF break and then compare with the LF obtained with the peak luminosity break obtained by other authors in the literature.

However, the values of these luminosity breaks are lower than those in \cite{pescalli2016A&A...587A..40P,Dong2022MNRAS.513.1078D,Yu2015ApJS..218...13Y,petrosian2015,Wanderman2015MNRAS.448.3026W,Wanderman2010MNRAS.406.1944W},
where the average log luminosity break is $51$.
This is because we use plateau luminosity for deriving the LF. 
The difference between the $\log$ flux of the prompt and plateau is of the order of $5$. This estimate is taken as the difference between an average of the logarithm of the peak flux of the prompt emission and the logarithm of flux at the end of the plateau emission in our sample. 
If we account for this in our luminosity break, then our values align with the literature.

\begin{figure}	
    \centering
    \includegraphics[width=\textwidth]{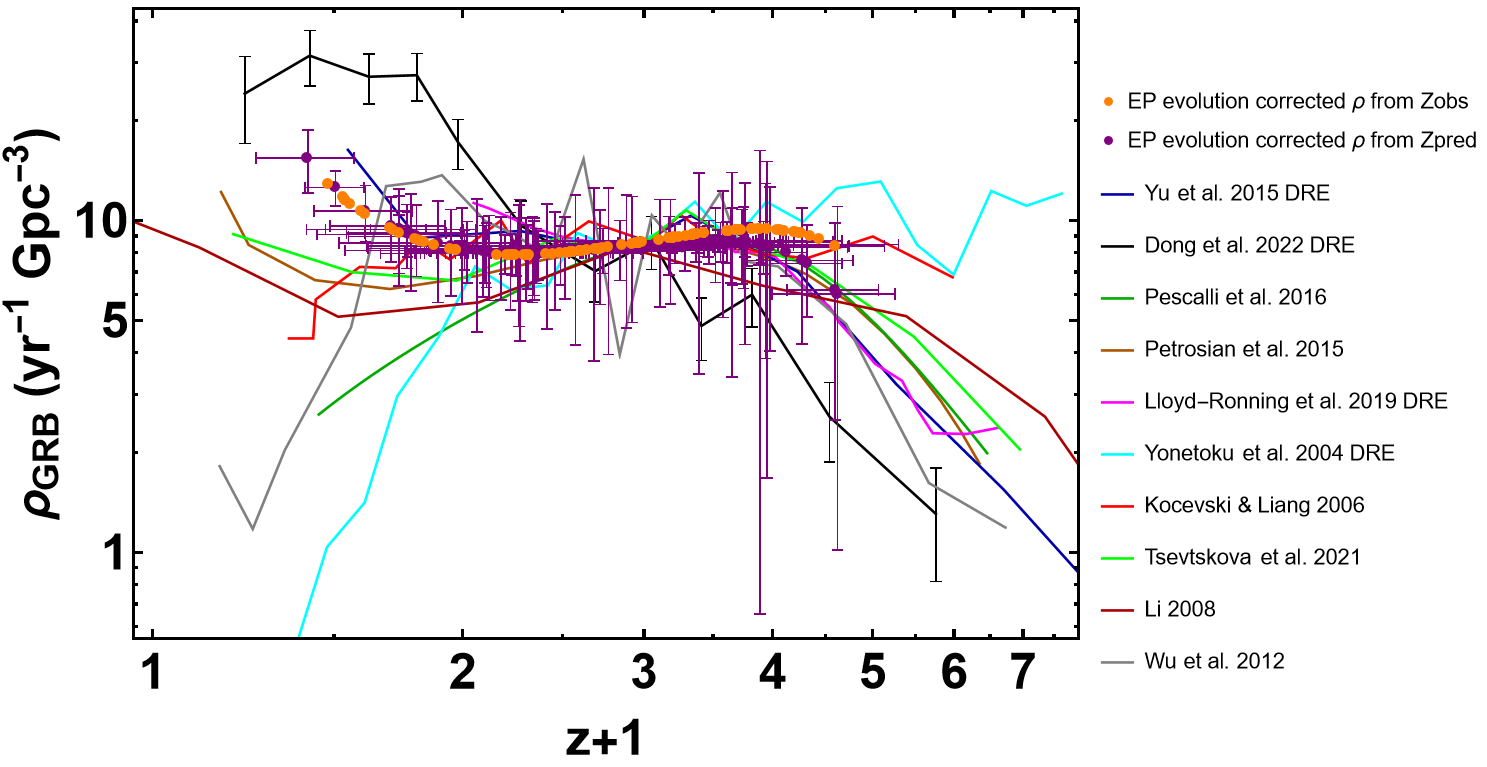}
    \includegraphics[width=\textwidth]{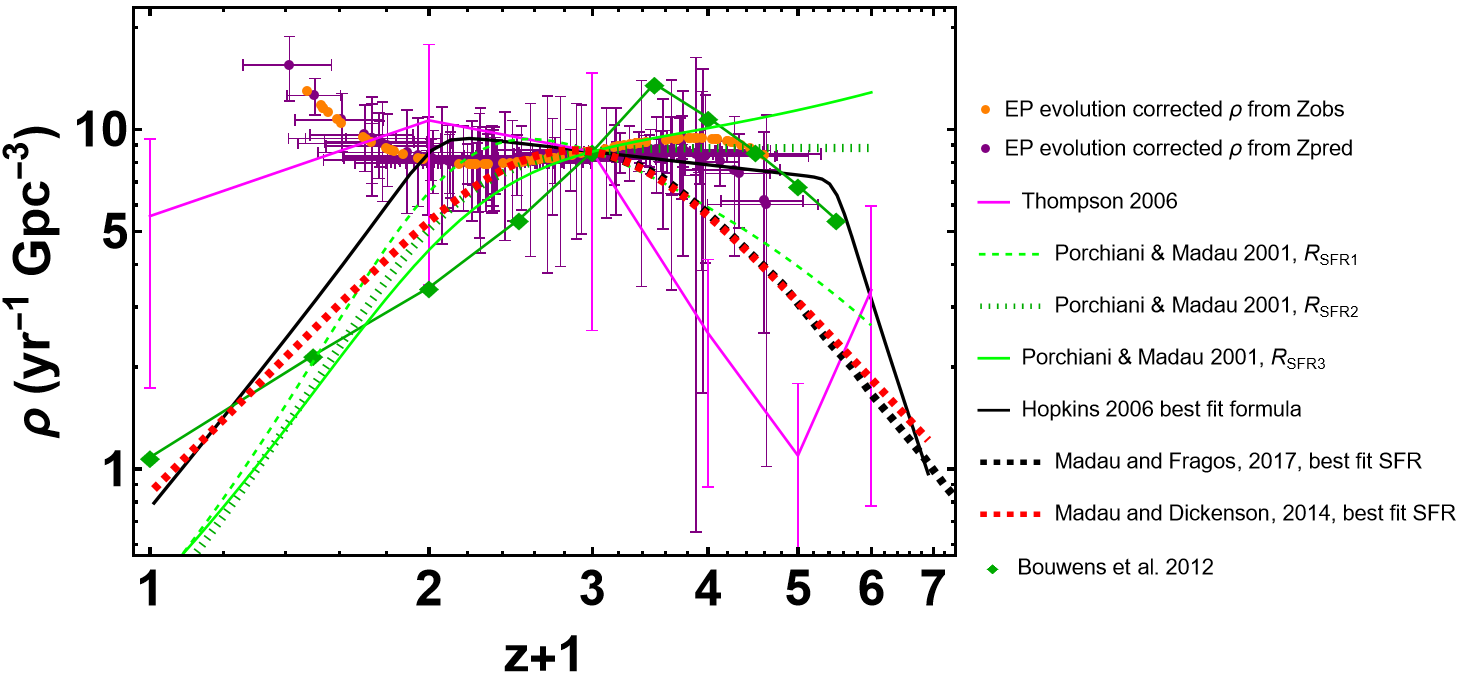}
    \caption{The orange and purple dots show the density rates obtained from $z_{obs}$ and $z_{pred}$, respectively. 
    Top Panel: the density rate evolution for our sample, compared with other published density rate evolutions for GRBs only. Bottom Panel: The density rate evolution for our GRB sample compared with the published SFR.
}
  \label{fig:densityrate}
\end{figure}

The density rates (see Fig. \ref{fig:densityrate}) are calculated using the same 97 GRBs as in the LF.
{The density rate are computed using the following formula:
\begin{equation}
{\dot\rho}(z)=Z[d{\dot\sigma}(z)/dz]/[dV(z)/dz]
\label{density}
\end{equation}

where Z = z+1, $dV(z)/dz$ is the comoving volume element, and $\dot\sigma$(z) is the cumulative of the number of GRBs as a function of redshift. 
We consider the years of observations of these 97 GRBs from 24th January 2002 until 4th February 2021 (19.04 years). 
Since GRBs in our catalog majorly come from Swift (88\%), then we also consider the field of view (FoV), and we remove the time in which Swift-BAT is not observing due to the South Atlantic Anomalies (SAA).
The FoV is 17\% of the whole sky, and the time of observations without the SAA for Swift-BAT is at 75.8\% \citep{Burns2016ApJ...818..110B}.
Furthermore, our sample GRBs have optical plateaus and constitute a 36\% fraction of the total sample. 
Thus, the density rate is adjusted to take this into account. }


The plots compare our density rate and those in the literature (upper panel of Fig. \ref{fig:densityrate}) and our density rate with SFR (lower panel of Fig. \ref{fig:densityrate}).
{We removed 7 GRBs which obtained a density rate going into the negative value within the error bars.
This is due to the large errorbars of the parameters involved.
Thus, the density rate plot features 90 GRBs as opposed to the 97 GRBs present in the luminosity function.}
We found the actual rate to be $\mathbf{8.47 - 9}$ Gpc $yr^{-1}$ for $1.9<z<2.3$ for the $z_{obs}$ and $z_{pred}$, respectively, without an arbitrary normalization. 
We have renormalized the rate (with a factor of 20) to account for our subset of LGRBs, compared to all LGRBs detected by Fermi and Swift, from the first detection of Fermi (14th July 2008) to 30th May 2023.
{We have quantitatively shown that the SFR does not match the GRB rate at low redshift ($z<1$). 
This discrepancy is of a factor of $\sim$3, comparing with \cite{Hopkins2006ApJ...651..142H}.
The SFR that better fit our GRB data is \cite{Hopkins2006ApJ...651..142H,Graham2016ApJ...823..154G}, where the agreement runs from $z=0.8$ until $z=4.8$.}

The GRB rate does not follow the SFR at $z<1$, supporting the previous studies with a quantitative estimate.
Although for most of the SFR compared here, the GRB rate does not follow the SFR at low-$z$, in the case of \cite{Thompson2006ApJ...647..787T}, there is no disagreement due to the very large errorbars.
In general, the nature of this discrepancy can be due to selection biases \citep{petrosian2015} or to metallicity \citep{Yoon2006A&A...460..199Y,Ghirlanda2022ApJ...932...10G}. 
The last two papers explain this excess due to the increased metallicity with cosmic time, which prevents a larger fraction of massive stars from producing GRBs. 
\cite{Ghirlanda2022ApJ...932...10G} found that a metallicity threshold of $12+log (O/ H) <= 8.6$ accounts for the derived GRB efficiency evolution.  
They verified their results, computing the stellar masses of galaxies with metallicities below the threshold value found and compared them with the observed masses of a sample of GRB hosts claimed to be complete. 
The maximum and average stellar masses are consistent with the measured masses of the hosts and their redshift evolutions. 
However, these results depend on the claim of the completeness of the sample of the GRB hosts, which can be debated.
Another relevant possibility is the low-$z$ GRBs have the same non-collapsar origin \citep{PetrosianDainotti2023arXiv230515081V} of the SGRBs. 
The resulting LF and density rate evolutions are consistent with the ones in the literature and are also consistent between the ones derived from $z_{obs}$ and $z_{pred}$. 
However, due to the large uncertainties in the measurements both interpretations of \cite{Yoon2006A&A...460..199Y} and \cite{PetrosianDainotti2023arXiv230515081V} are viable. 
We have shown in the right panel of Fig. \ref{fig:lf} that adding the estimated redshift improves the derivation of the break of the LF by 90\%, namely $46.6\pm0.3$ vs $47.2\pm0.03$.

\section{Summary and Conclusion}\label{summary}

We tested an ensemble with several algorithms to determine the redshift of GRBs based on characteristics related to their prompt and, for the first time, the plateau emission. 
Some afterglow characteristics (duration of the plateau emission, its flux, and the decay after the plateau) are among the most powerful predictors.
A successful application of these methods to the Swift existing GRB catalog can nearly double the sample of GRBs with observed redshift and observed plateau, thus yielding a much larger sample. 
This, in turn, has allowed us to determine the LF and the density rate evolution of GRBs with unprecedented accuracy.
Using a sample of 147 GRBs from the Swift satellite with known redshifts, we demonstrate the SuperLearner's capabilities to provide accurate $z_{pred}$ that agree with the $z_{obs}$ with a Pearson correlation coefficient of $r=0.93$, and MSE=$0.21$. 
This result makes GRBs powerful distance estimators, improving upon other existing methods. 
Indeed, we obtained an increase of $61\%$ in the Pearson correlation of the $z_{pred}$ and $z_{pred}$. 
Instead, previous works, which omitted plateau parameters and used only one method, achieved a correlation of $r=0.57$. 
Results from applying this method to a large number of LGRBs have cast light on the controversy on the discrepancy between the density rate of LGRBs and the cosmological SFR and to accurately measure the GRB LFs. 
Our results mirror similar deviations in the formation rates of LGRBs found by \cite{Tsvetkova2021ApJ...908...83T,Lloyd-Ronning2019MNRAS.488.5823L,Yu2015ApJS..218...13Y,petrosian2015,pescalli2016A&A...587A..40P}.
Notably, \cite{pescalli2016A&A...587A..40P} suggests a deviation of 2.6 sigma when considering the star formation history of \cite{madau2014ARA&A..52..415M}. 
Furthermore, \cite{Dainotti2021ApJ...914L..40D} also arrived at a similar conclusion regarding SGRBs at low redshifts, even though the delayed SFR has been accounted for.
Our primary conclusion, which aligns with findings reached by \cite{PetrosianDainotti2023arXiv230515081V}, is that the population of LGRBs at low redshift includes a component of compact merger progenitors, similar to the SGRBs. 
A different explanation for the excess of GRBs at low redshift, originally proposed by \cite{Yoon2006A&A...460..199Y}, is related to the increase in metallicity with cosmic time, which prevents a larger fraction of massive stars from producing GRBs.


\section{Acknowledgement}
We are particular grateful to Michal Ostrowski for his precious discussion during the 
data analysis and results, to Trevor Nelson for his preliminary analysis during the summer internship program at Stanford University supported by SLAC, to Simona Scimemi for the preliminary analysis related only to prompt emission variables during the winter internship 2018 at INAF Bologna, to Monica Brora on the very preliminary analysis on random forest. 
We also grateful to Macjei Bilicki for the useful discussion on the parameter space division. We thank particularly Douglas Figberg for his interesting discussion on selection biases related to this procedure and the way on how to solve them. M.G.D is particularly grateful for the support received as RIKEN pionieer research during her stay in Riken. M.G.D is grateful to the Gill Ferrandes and Donald Warren for their interesting discussions on the method of SuperLearner. 
We are also grateful to Trevor Hastie for useful suggestions to avoid overfitting and piece of the code for the 10-fold cross-validation. 
We are also grateful for the initial discussion on the project to Dmitry Maleshev. M.G.D. acknowledges the funding from MINIATURA2, Numer: 2018/02/X/ST9/03673.
M.G.D also acknowledges Prof. Peter Behroozi for the insightful discussion on how to avoid overfitting in our statistical learning models.
M.G.D and A.N also thank Nikita Khatiya, Dhruv Bal and Dr. Dieter Hartmann for extremely valuable feedback on the DRE analysis.
This research was also support by the Visibility and Mobility module of the Jagiellonian University (Grant number: WSPR.WSDNSP.2.5.2022.5) and the NAWA STER Mobility Grant (Number: PPI/STE/2020/1/00029/U/00001).
AN would also like to acknowledge the NAOJ Exploratory Grant and Division of Science for supporting his stay in Tokyo.
We acknowledge funding from Swift GI Cycle XIX, grant number 22-SWIFT22-0032.
This work was also supported by the Polish National Science Centre grant UMO-2018/30/M/ST9/00757 and by Polish Ministry of Science and Higher Education grant DIR/WK/2018/12.
Numerical computations were in part carried out on Small Parallel Computers at the Center for Computational Astrophysics, National Astronomical Observatory of Japan.

\newpage

\appendix

\section{Methods}\label{Methods}
\subsection{Statistical learning methods}\label{ML}
We have used predictive data mining, also called supervised learning. It is based on prior knowledge of a ‘training’ data set on which we can build models to predict the relations in a new “test” data set. Examples are regression tools, such decision trees, random forests, and gradient boosting. These allow us to estimate complicated non-linear relationships between the response and {predictors} as well as high-order interactions between predictors. Such non-parametric methods are very powerful when the data set contains many observations.
{However, they suffer from the so-called “curse of dimensionality”, which limits the number of {predictors} one can efficiently estimate for a small sample size. 
As a result, the fully non-parametric machine learning methods allow using only a limited number of predictors, e.g., when estimating redshifts based on small GRB training sets. On the other hand, when important {predictors} are identified by other methods or expert knowledge, these methods allow the construction of efficient predictors without assumptions on the underlying form of the relevant correlations among the variables. 
In our analysis, we have used supervised {statistical learning tools} that employ regression. 
Within regression methods, we have used parametric, semi-parametric, and non-parametric approaches.
We give a brief description of each model below.}

\subsubsection{Predictors selection}
\begin{figure}[h!]
    \centering
    \includegraphics[width=0.5\textwidth]{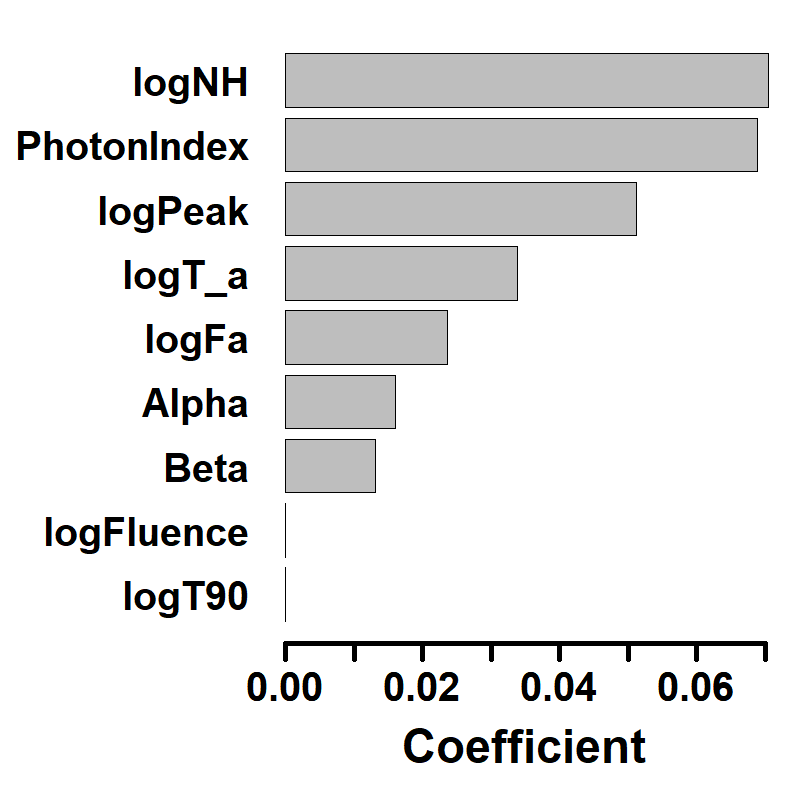}
    \caption{LASSO-based {predictors}  selection, showing the weights assigned to the different variables.}
    \label{fig:lassoF}
\end{figure}

The GRB sample contains a total of nine predictors. 
Four belong to the prompt:
$Fluence$, $T_{90}$, $\log(Peakflux)$, and $Photon$ $Index$. 
The rest belong to the plateau and afterglow: $T_a$, $F_a$, $\alpha$, $\beta$, and $NH$.
To pick out the best predictors of redshift among these 9 variables, we applied the Least Absolute Shrinkage and Selection Operator (LASSO) {predictor} selection technique.
The LASSO technique applies a shrinkage method for linear regression by forcing the $\ell^1$ norm (sum of the magnitude of all vectors in the given space)
of the result vector to be $\le \lambda$, where $\lambda$ is a positive number known as the tuning parameter.
This form of penalization enables the model to choose specific {predictors} while nullifying others by assigning their coefficients a value of 0 \citep{TibshiraniLasso}.
The tuning parameter determines the degree of shrinkage applied to the estimated vector. 
Consequently, when using fewer {predictors}, the model becomes more interpretable and generally exhibits lower prediction errors than the full model.
For our analysis, we use the GLMNET function with the LASSO selection {predictors} \citep{hastie2017extended,tibshirani2012strong}.
We pick the $\lambda.1se$ value, the maximum $\lambda$ value for which the error is within one standard deviation \citep{hastieTibs}, and its corresponding coefficients for the predictors.
Thus, we drop $T_{90}$ and $Fluence$ while keeping the rest of the 7 variables.
The coefficients assigned by LASSO to each of them are displayed in Fig. \ref{fig:lassoF}, and we choose only the non-zero coefficient {predictors}.
The LASSO selected {predictors} are also presented in Fig. \ref{scatter}. 

\subsubsection{The Generalized Additive Model}
In the {semi-parametric generalized additive models (GAM)}, $Y$ is related to the predictors,
$x_1,\ldots,x_p$ through the additive model $f(x_1,\ldots,x_p)=\beta_0+\sum_{i=1}^{p} {\beta_i
f_i(x_i)}$. 
The functions $f$ may have some specified parametric form (e.g., a polynomial) or be
estimated non$-$parametrically, simply as `smooth functions.' In this method, interactions between different predictors can be included. 
An interaction may arise when there are two or more {predictors}, and the influence of some of them on the response variable depends on the other {predictors}. 
In practice, interactions make it more difficult to predict the consequences of changing the value of a variable. 
If any smooth functions were allowed in model fitting, then the maximum likelihood estimation of such models would result in over-fitting. 
Thus, the models are usually fit by penalized likelihood maximization, in which the model is modified by adding a penalized likelihood estimation, namely a penalty for each smooth
function, penalizing its irregularities. 

We use GAM to determine an analytical formula that best estimates the $\log(z+1)$ value.
It should be noted that we present results in log scale because of the log scale transformations we applied to the majority of the variables and because the best model is indeed trained in $\log(1+z)$. 
In addition, Z = $z+1$ is a more natural parametrization of the cosmological evolution.
The parameters of this formula are the LASSO selected predictors and their cross multiplicative terms (terms defined as the products among the 7 chosen variables). 
A total of 16,511 formulas were generated considering all possible combinations of all seven {predictors} and the product of the pairs of these variables.

\begin{figure}
    \centering
    \includegraphics[width=0.49\textwidth]{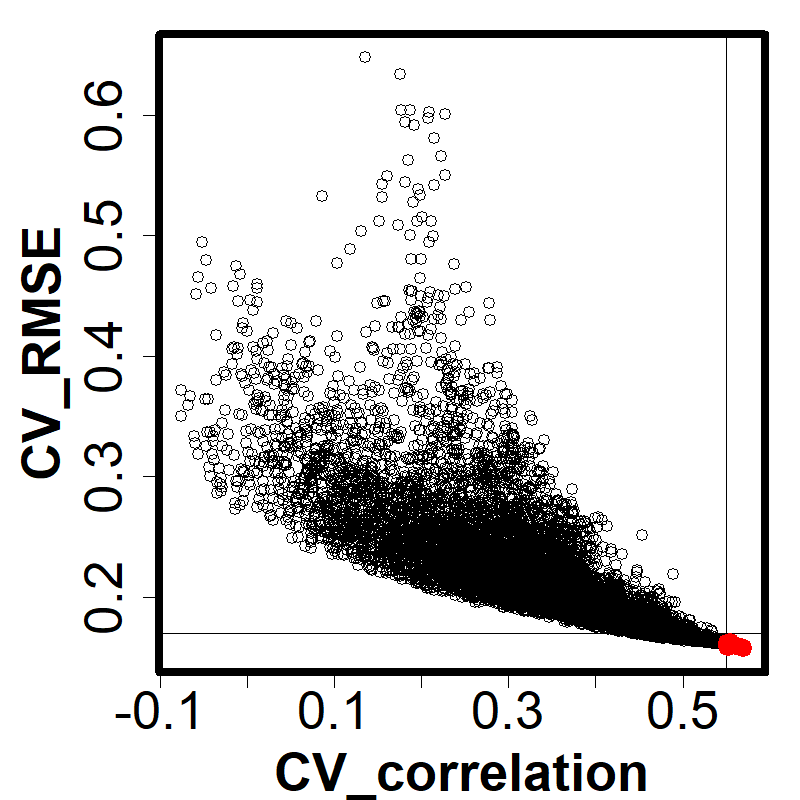}
    \includegraphics[width=0.49\textwidth]{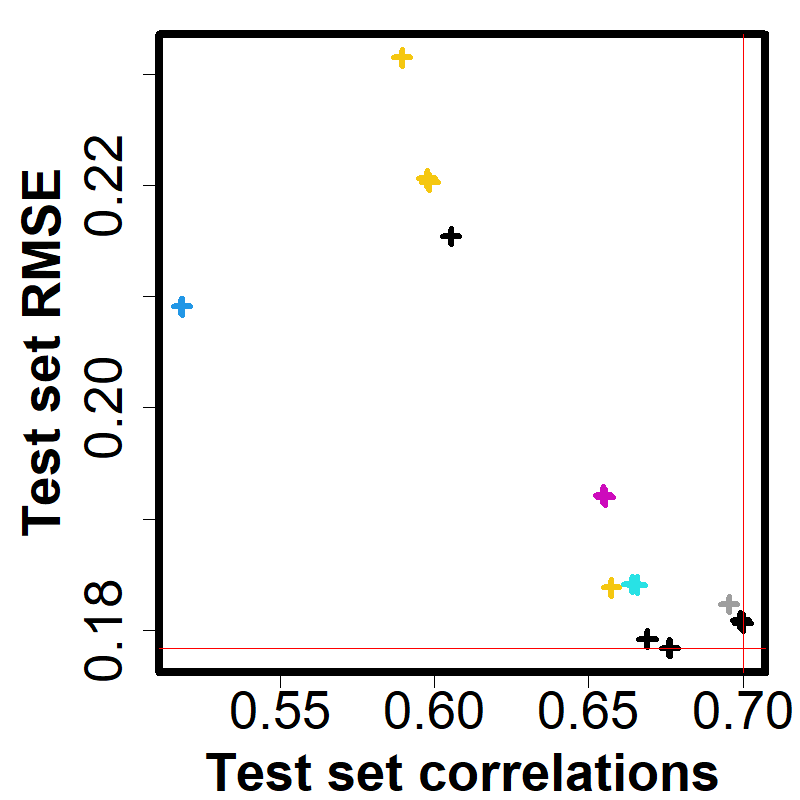}
    \caption{
    Left panel: The cross-validated RMSE-vs. the correlation coefficient between $z_{obs}$ and $z_{pred}$ for 16,511 formulas. 
    The red points show the formulas selected based on the correlation and RMSE cut-off (shown by the black lines).
    Right panel: The same results of the left panel but on a separate and unseen test set.
    The red lines show the highest correlation, and lowest RMSE obtained on test set.
    }
    \label{fig:gam_formula}
\end{figure}

To guarantee that this extensive search among all the possible combinations, the 16,511 GAM models, of the seven variables is reliable, we have divided our sample into a training set (90\% of the total sample) and a test set (the 10\%). 
The extensive search is performed only on the training sample and tested on the test set. 
Then, we tested all possible models and have chosen only the ones for which $r \geq 0.55$ and RMSE  $\leq 0.22$ to be validated, see left panel of Fig \ref{fig:gam_formula}. 
Then, we choose among these most performant models (denoted with red dots in the figures) the ones with the highest Pearson correlation between $\log(z_{obs}+1)$ and $\log(z_{pred}+1)$, the lowest root mean square error (RMSE, the square root of the mean squared error $<\Delta z^2>$), and the lowest median absolute deviation (MAD, median value of the absolute values of $\Delta z$) on the test set, see Fig. \ref{fig:gam_formula}. 
All the GAM models were fit to our data set using GAM.
Among those, we selected the three most performant formulas, as stated above. 

\begin{dmath}
\log(z+1) = (\log(Peak)^2 + \log(NH)^2 + PI^2 + \beta^2 + \log(Peak) + \alpha)^2 + \log(F_a) + \log(T_a) + \beta + PI + \log(NH) + \log(F_a)^2 + \log(T_a)^2 + \alpha^2
\label{bestcorrelation}
\end{dmath}

\begin{dmath}
\log(z+1) = (\log(Peak)^2 + \log(NH)^2 + \beta^2 + \log(Peak) + \alpha)^2 + \log(F_a) + \log(T_a) + \beta + PI + \log(NH) + \log(F_a)^2 + \log(T_a)^2 + \alpha^2 + PI^2
\label{bestRMSE}
\end{dmath}

\begin{dmath}
\log(z+1) = (\log(Peak)^2 + \log(NH)^2 + \beta^2 + \log(Peak) + \log(NH))^2 + \log(F_a) + \log(T_a) + \alpha + \beta + PI + \log(F_a)^2 + \log(T_a)^2 + \alpha^2 + PI^2
\label{bestMAD}
\end{dmath}

\noindent
where $PI$ is the prompt $Photon$ $Index$, $Peak$ is the prompt $Peak flux$, and the remaining parameters are as defined previously in Sec. \ref{sample}.
To test the large number of formulas, we leveraged the More high-compute cluster of the Center for Computational Astrophysics at the National Astronomical Observatory of Japan.

\subsubsection{Step-AIC}
{Step-AIC} is a parametric approach that chooses algorithms based on the Akaike Information Criterion (AIC) \cite{sakamoto1986akaike,Venables02statisticscomplements}, which, given a collection of models, estimates the quality of each model relative to each other. The StepAIC model works by fitting a formula to which it adds (or subtracts) {predictors} iteratively while minimizing the AIC.

\subsubsection{The Generalized Linear Model}
{Generalized Linear Model} (GLM) \cite{glm} is a parametric method that estimates the mean of the response variable's distribution using a linear model of the {predictors}. 
The response variable is connected to the desired target variable (GRB redshift) via a link function. 
This is an extension of the ordinary linear model. 
Using the appropriate link functions, we can consider non-Gaussian target variable distributions, such as Poisson or Gamma functions.
GLM adopts the iteratively weighted least squares method for maximum likelihood estimation to fit these functions.
Here, we assign all three formulas (Eq. \ref{bestcorrelation}, \ref{bestRMSE}, and \ref{bestMAD}) as the desired fitting functions, while the link function is Gaussian.

\subsubsection{The Bayesian Generalized Linear Model}
{Bayesian Generalized Linear Model} (BGLM) is a Bayesian inference of the generalized linear model. It determines the most likely estimate of the response variable and the prior distribution on the set of regression parameters (Maximum A Posteriori estimator, MAP) given the particular set of predictors.
BGLM tries to estimate the most likely value of the unknown parameter that can generate the observed data. This is known as the Fisher principle. It is numerically and computationally stable.
A BGLM implements a student-T prior distribution for the regression coefficients.
Then, given the observed data, the likelihood function for these parameters is calculated. The likelihood function and priors are combined to produce the posterior distributions from which we obtain the MAP estimators of the desired parameters 
\citep{birnbaum1962foundations,hastie1987generalized,hastie1990generalized,friedman2010regularization}.

\subsection{Outlier removal with M-estimator}
To remove outliers that might not represent the full sample, we apply the Robust Linear regression Method (RLM) with the {M-estimator}.
This is an alternative technique to the ordinary least squares method, which
for a univariate {response variable}, $Y$ and a vector $X=(X_1, \ldots, X_p)$ of {predictors}
$X$, attempts to fit the following relation: $f(x_1,\ldots,x_p)=\beta_0+\sum_{i=1}^{p}{\beta_ix_i}$.
The ordinary least squares method attempts to minimize the square of the residuals (called the $L_2$ norm regression). 
However, this gives outliers of the data set a higher weight, which significantly affects the results of the regression fit.
In contrast, the M-estimator attempts to minimize the sum of a function of the residuals. 
The function chosen for our analysis is the Huber Function \citep{huber1964}.

Iterated re-weighted least squares do the M-estimator fitting, and the weights, ranging from 0 to 1, are assigned based on the residuals \citep{huber1996robust}.
RLM is used for the detection of highly influential observations.
We are using the implementation of RLM as described in the MASS package of R \citep{venables2002random}.
{We call \textit{rlm()} function and specify the Eq. \ref{bestcorrelation} based on which the fitting is performed. 
To ensure the \textit{rlm()} function uses the M-estimator method for assigning weights, we specify the parameter \textit{method=`M'}, where M indicates the M-estimator techniques.
For further discussion on robust statistical methods see \cite{stromberg2004write,hampel2011robust,marazzi1993algorithms}.
}

Here, choose a cut-off weight of 0.5, eliminating GRBs with a weight $\leq 0.5$.
The threshold of 0.5 was chosen to ensure we remove less than 5\% of our sample as outliers.
With the M-estimator, the sample size is reduced from 150 to 147.

\subsection{The SuperLearner and the Relative importance of the {predictors}}\label{SuperLearner}
{SuperLearner} is an algorithm that uses internal 10-fold cross-validation (10fCv) to estimate the performance of multiple {statistical learning models} or the same model with different settings \citep{van2007super,naimi2018stacked}. 
It minimizes the estimated risk (a measure of model accuracy) by minimizing the RMSE. 
To test the performance of our results and estimate the prediction error, besides the 10fCV, we also used an external layer of cross-validation called nested cross-validation.
Namely, we performed 10fCV repeatedly $100$ times, with the folds of the cross-validation step being randomized for each iteration. 
This type of cross-validation involves partitioning the sample into complementary subsets (in this case, $10$), fitting the model on 9 out of $10$ folds (training sets) of the full data set, and validating the analysis on the remaining 1 testing set. 
The procedure is repeated iteratively so that each set of the $10$ will be used as a test set and the remaining as training sets, and the prediction results are averaged over the number of runs (in this case $100$). 
This step allows us to remove the randomness inherent to the cross-validation procedure and stabilize the results.
Then, SuperLearner creates an optimal weighted average of those models, e.g., an ensemble, using the test data performance. 
Namely, the SuperLearner provides coefficients to inform of each learner's weight, $A_i$, or importance in the overall ensemble. 
By default, the weights are always greater than or equal to $0$ and sum to $1$. 
This approach is asymptotically as accurate as the best possible prediction algorithm tested. 
Our results show that the model produced by BGLM using Eq. \ref{bestMAD} obtains the highest weight ($A_1=0.46$), followed by StepAIC ($A_2=0.25$), GLM using Eq. \ref{bestcorrelation} ($A_3=0.16$), GAMs using Eq. \ref{bestcorrelation} ($A_4=0.06$), Eq. \ref{bestMAD} ($A_5=0.02$) and Eq. \ref{bestRMSE} ($A_6=0.01$).
The rest of the models have a non-zero weight $>1\%$.

These coefficients inform us which is the best model and reflects the weighted average of multiple models.
The metric we use to define the goodness of our results is based on the minimization of the RMSE in the SuperLearner algorithm.
To further measure the goodness of our results, along with the SuperLearner RMSE, we use the following metrics:
the Pearson correlation coefficient, $r$, between $z_{obs}$ and $z_{pred}$, the mean squared error, MSE, and the bias.

\begin{figure}[H]
    \centering
    \includegraphics[width = 0.49\textwidth]{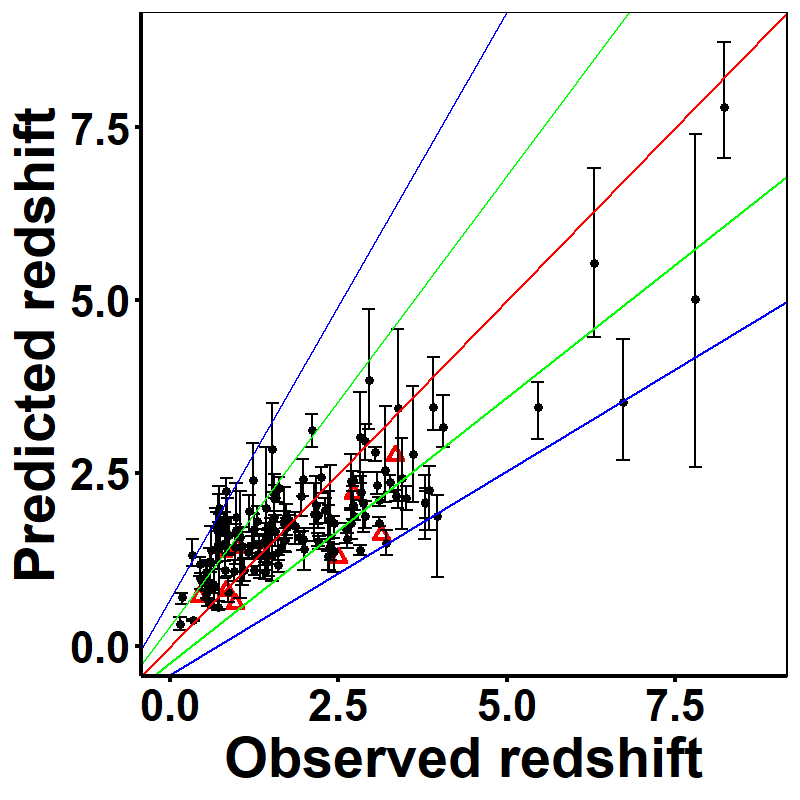}
    \includegraphics[width = 0.49\textwidth]{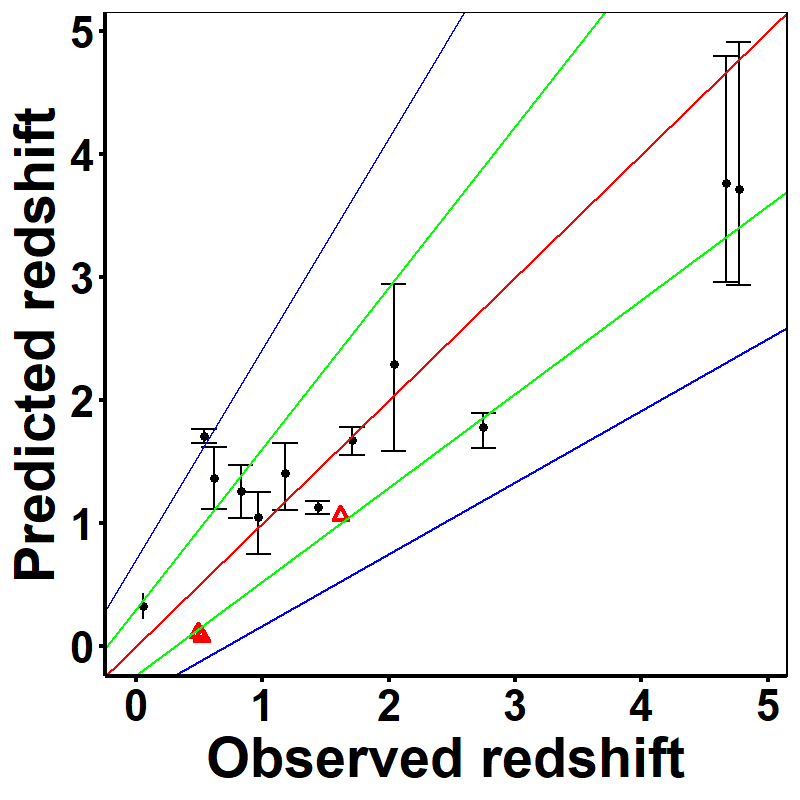}
    \caption{Left panel: The $z_{obs}$ vs $z_{pred}$ scatter plot for a training set of 147 GRBs. 
    Right panel: The $z_{obs}$ vs $z_{pred}$ scatter plot for a test set of 17 GRBs.}
    \label{fig:traintest}
\end{figure}

We can see that both results are a good reproduction of the observed redshifts for the training set (see the left panel of Fig. \ref{fig:traintest}) and a test set of 17 GRBs (see the right panel of Fig. \ref{fig:traintest}).
We stress that the goodness of the initial results is due to the procedure of Superlearner. 
{More specifically, for the training set, we obtained $RMSE=0.83$, $r=0.82$ and the bias$=0.14$, 
for the test set $RMSE=0.62$, $r=0.91$ and the bias $=0.1$ as we can see from the left and right panels of Fig. \ref{fig:traintest}, respectively.}
It is clear that we consistently obtain high values of $r$ and low values of RMSE, and the bias for both samples indicates that the ensemble model is not overfitting. 

Our results show that the accuracy of the prediction is stable for the majority of partitions. 
However, we observe a higher RMSE (0.62) and a slightly smaller correlation coefficient of 0.91 for a small number of partitions. 
This is indeed natural due to the large heterogeneity of the data and the relatively small sample size. 
{It should be noted that the default SuperLearner implementation of the methods mentioned above, namely GAM, GLM, and BGLM, utilizes a default formula for fitting the data, a smoothed linear combination of the {predictors}.
We have edited the definitions of these functions in R to ensure they use Eq. \ref{bestcorrelation}, \ref{bestRMSE}, and \ref{bestMAD} as the fitting functions.}

To estimate the contribution of each predictor, we use the relative importance of {predictors}, see the right panel of Fig. \ref{influence}. This is an average of local importance given by a local linear approximation of prediction. 
Specifically, we create corresponding synthetic data for every observation by adding Gaussian noise. 
Next, we construct an approximate change in prediction through a linear model prediction, P= X$B$ where $B$ are fit coefficients, and we define the local relative importance of feature $i$ over all sample points by $R_i =|B_i|/\sum|B_i|$. 

\subsubsection{Errorbars on predicted redshift and the catastrophic outliers}\label{errorbarSection}

For the 1$\sigma$ and 2$\sigma$ cones shown in the left panel of Fig. \ref{influence}, we first obtained the residuals between the $\log(z_{pred}+1)$ and $\log(z_{obs}+1)$. 
Then, we fit this residual distribution with a Gaussian fit and derive the standard deviation values of this fit. 
In the linear scale, the cones are obtained by:
$$
1\sigma = 10^{\sigma*}z + (10^{\sigma*}-1)
$$
$$
2\sigma = 10^{2\sigma*}z + (10^{2\sigma*}-1)
$$
where $\sigma*$ is the standard deviation of the residual Gaussian fit obtained in the $\log(z+1)$ scale.

The GRBs in our sample have an uncertainty measurement for $F_a$, $T_a$, $\alpha$, $Peak flux$, $Photon$ $Index$, and $\beta$. 
The uncertainties on the redshift are not available, and thus, they are not used. 
To incorporate these uncertainty measurements into our prediction, we perform Monte Carlo simulations, assuming a Gaussian distribution for the uncertainties of each variable. 
This assumption is indeed reasonable because the uncertainties on the variables are independent between measurements of the same variable, which is random. 
During the 10fCV, for the GRBs present in the test fold, we produce 100 simulated GRBs from the simulation and predict their redshifts. 
This provides us with a maximum and minimum uncertainty on the $z_{pred}$ for each GRB, which considers the measurement uncertainties on the variables. 
The errorbars presented in the correlation plot in the left panel of Fig.\ref{influence}, both panels of Fig. \ref{fig:traintest} and Fig. \ref{results_10fcv}, are generated in this manner.
We have removed the so-called catastrophic outliers.
They are defined in \cite{jones2020tests} as the GRBs for which $|{\Delta z}| > 2\sigma$, and thus they lie outside the blue cone presented in Fig. \ref{results_10fcv}. 
In our analysis, we consider both all data (left panel of Fig. \ref{results_10fcv}) or data for which the catastrophic outliers have been removed (the left panel of Fig. \ref{results_10fcv}).
We remove the errorbars from the plots for the cases in which $\delta z_{pred}/z_{pred} \geq 1$ and show them as red triangles. 

\subsection{Optimal Transport bias correction}
\begin{figure}[H]
\includegraphics[width = 0.49\textwidth]{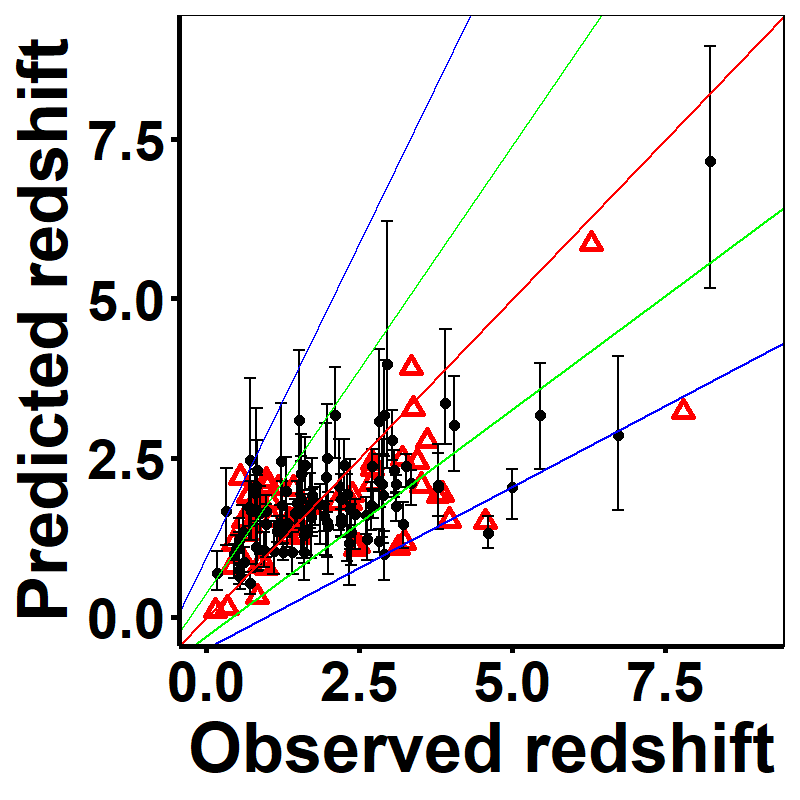}
\includegraphics[width = 0.49\textwidth]{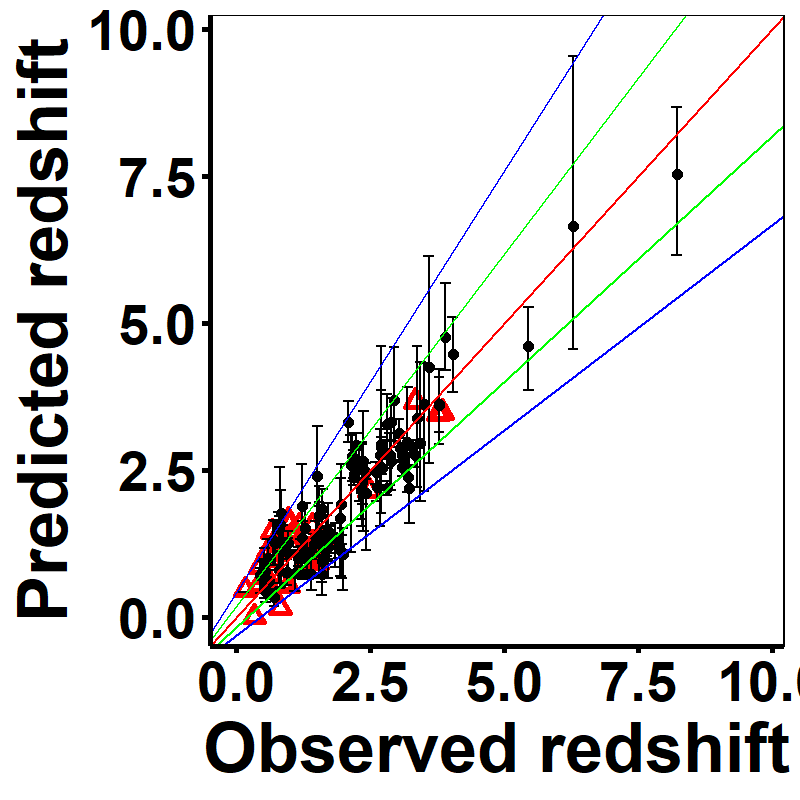}
\caption{
{The scatter plot between $z_{obs}$ and $z_{pred}$.}
Left panel: Predictions of 147 GRBs, before removal of catastrophic outliers and bias correction
Right panel: The correlation for 136 GRBs after removal of catastrophic outliers and application of bias correction. 
}
\label{results_10fcv}
\end{figure}

When the predictions of a {statistical learning model} are skewed towards a particular range of values, then such a model is said to be biased.
Bias correction techniques attempt to minimize such biases in the model predictions.
Bias is defined as the mean of the difference between the model's prediction and the actual values.
Biased results from {statistical learning} models generally arise due to imbalanced training data, where more observations in a particular range are present compared to others.
In the case of GRBs, there are more observations at low redshifts compared to high redshift ones.
Thus, a {statistical learning} model trained on such data will be more inclined toward underestimating the redshift of GRBs.
\newline\indent
To correct this underestimation, we apply the bias correction technique of Optimal Transport.
In this technique, a linear model is fit to the sorted values of $\log(z_{obs}+1)$ and $\log(z_{pred}+1)$. 
Then, the corrected predictions are obtained as shown:
$C_{pred} = A \times U_{pred} + B$, where $A$ and $B$ are the slope and intercept of the linear fit, respectively, and $C_{pred}$ and $U_{pred}$ are the corrected and predicted $\log(z+1)$, respectively.
\newline\indent
Here, the above-described bias correction method is applied independently in three regions{:} $z_{obs}<2$, $2<z_{obs}<3.5$, and $z_{obs}>3.5$.
This enables us to carefully eliminate the bias in each section of the response variable, providing us with more accurate results.
\newline\indent
It should be noted here that the bias correction is applied following the removal of the catastrophic outliers, specifically the points outside of the blue cone in the left panel of Fig. \ref{results_10fcv}. 
{We apply the bias correction after the catastrophic outliers removal, because we need to guarantee that the prediction obtained by our training sample is not initially biased by the catastrophic outliers. This will allow us to train in a parameter space that is driven by predictors in the 2$\sigma$ cone and not by predictors outside, thus resulting in a more reliable prediction.}

The right panel shows the prediction after removing the catastrophic outliers and the bias correction.
We also stress that if we remove the catastrophic outliers additionally after the bias correction, we obtain an even higher correlation coefficient between the $z_{pred}$ and $z_{obs}$ of 0.94.
The bias correction can also be applied in the generalization sample as long as 
the training set and the generalization sample are drawn by the same parent population.

\subsection{EP method}\label{EPmethod}
We here describe mainly the Efron and Petrosian method. 
Indeed, there are various {\it non-parametric, non-binning methods} that are more powerful, especially for small samples. 
For example, \cite{Schmidt1968ApJ...151..393S}, using the so-called $V/V_{\rm max}$ method on a sample of 33 quasi-stellar radio sources, with one-sided truncation (due to a lower limit on the flux) was able to determine their LF and density evolution.
A review by Petrosian (1992) describes the further development of this and other similar methods with the conclusion that all these methods are equivalent to the more general \cite{Lynden1971MNRAS.155...95L} $C^{-}$ method. 
However, as also pointed out in the Petrosian review, all these methods require the
critical assumption that the variables are uncorrelated, or independent.
This shortcoming led to the development of the more powerful (also non-parametric, non-binning) methods by \cite{Efron1992} (EP),
which first tests for correlation between the variables. 
If such correlations are found, the EP method uses a modified Kendall-Tau test to yield an uncorrelated set of variables.
It then proceeds with determining the distribution of the uncorrelated variables using the $C^{-}$ or other non-parametric method.
This approach was later generalized to two-sided truncated data (e.g.~data with upper and lower flux limits) in Efron \& Petrosian (1999). 
This combined Efron-Petrosian and Lynden-Bell method (EPL) has been proven to be very useful for studies of many aspects of Active Galactic Nuclei 
(see, e.g., \cite{Maloney1999ApJ...518...32M,Singal2011ApJ...743..104S,Singal2013HEAD...1330007S}
and LGRBs
(\cite{Lloyd1999ApJ...511..550L,Lloyd2000ApJ...534..227L,Lloyd2002ApJ...574..554L,Kocevski2006ApJ...642..371K,yonetoku2004,Yonetoku2014ApJ...789...65Y,dainotti2013slope,dainotti2015,dainotti2017a,petrosian2015,Yu2015ApJS..218...13Y,pescalli2016A&A...587A..40P,tsvetkova2017konus}.
More specifically, the EP method uses a modified version of the Kendall $\tau$ statistic (a rank measure) to test the independence of variables in truncated data.
The rank of each data point is determined from amongst its ``associated sets", which include all objects that could have been observed given the observational limits. This method uses the Kendall rank test to infer the best-fit values of parameters that determine the evolution of the LF, see \cite{Dainotti2013a, petrosian2015} for additional details.

\begin{figure}
    \includegraphics[width=0.49\textwidth]{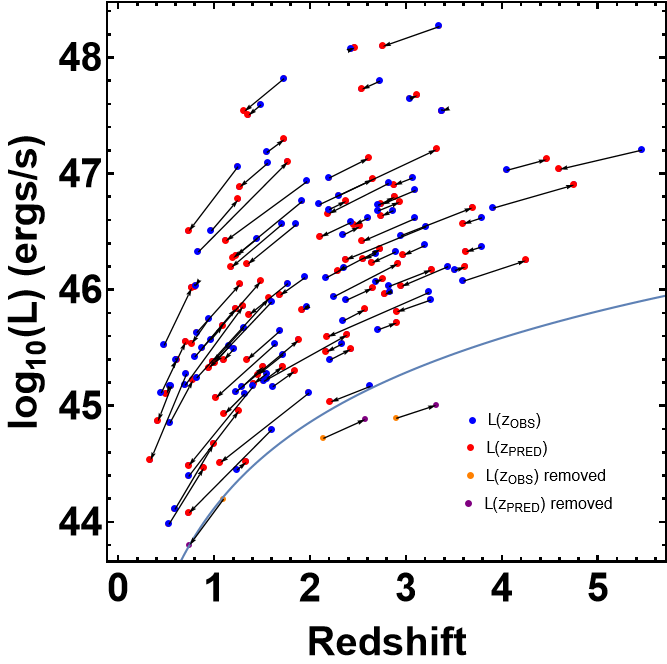}
    \includegraphics[width=0.49\textwidth]{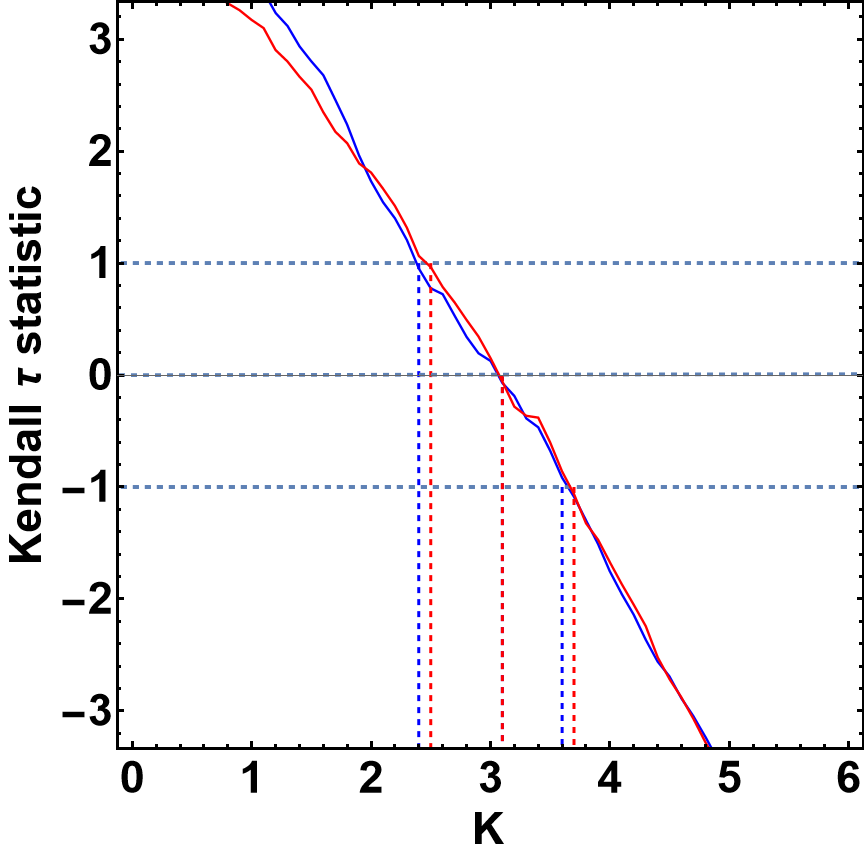}
    \caption{
Left Panel: Luminosity vs. redshift for 103 GRBs. 
The luminosities calculated using $z_{obs}$ and $z_{pred}$ are shown with blue and red points, respectively.
The orange and purple points represent the luminosities of $z_{obs}$ and $z_{pred}$, respectively that are lower than the luminosity threshold given by the flux limit of $1.25\times10^{-14}erg/cm^{2}$. 
The arrows connect a GRB's $z_{obs}$ and luminosity value with its predicted values.
Right panel: $\tau$ versus $k$, where $k$ defines the power of the $g(z)$ function for the luminosity evolution. 
The red curve shows results for the GRBs with observed $z_{obs}$, while the blue curve yields the corresponding function using $z_{pred}$.
The $\tau=0$ values are obtained at {$k=3.07$} for the GRBs with {$z_{obs}$} and $3.06$ for {$z_{pred}$.}
}
\label{luminosities}
\end{figure}

The EP method requires a limiting flux, $F_{lim}$. 
For $F_{lim}$ we chose $L_{lim}= 4 \pi D_L^2(z) \, F_{lim} K$, where $D_L$ is the luminosity distance and $K$ is the K-correction for cosmic expansion \citep{bloom2001}, which gives the minimum observed luminosity for a given redshift. 
In the left panel of Fig. \ref{luminosities}, we show the limiting luminosity for $K=1$, not to have fuzzy boundaries, but for an appropriate evaluation of the luminosity evolution, we assign to each GRB its own K correction. 
We have investigated several limiting fluxes to determine a good representative value while keeping an adequate sample size. 
We have finally chosen the limiting flux $F_{lim} = 1.25\times10^{-14}erg/cm^{-2}$, for both the observed and predicted redshift, which allows $97$ GRBs. 
Limiting luminosity and luminosity of GRBs from the observed (red circles) and predicted redshifts (blue circles) are shown on the left panel of Fig. \ref{luminosities}. 
The orange and purple circles show the removed GRBs based on the chosen flux limit.

To estimate the luminosity evolution and density rate evolution of our samples, it is necessary first to determine whether the variable $L_{plateau}$ is correlated with redshift (i.e. if there is luminosity evolution).
If $L_{plateau}$ is statistically independent of $z$, a lack of this correlation implies the absence of such an evolution.

We determine $g(z)$, which is the function that describes the evolution, by defining a new variable $L'_{plateau} \equiv L_{plateau}/g(z)$ and obtaining the parameter that characterizes $g(z)$ so that $L_{plateau}$ is no longer correlated with $z$. 
The evolutionary function is commonly parameterized by a single parameter $g(z)=(1+z)^k$, see Sec. \ref{EPmethod}. We use this function for both the $z_{obs}$ and $z_{pred}$.
In the right panel of Fig \ref{luminosities}, we show the test statistic $\tau$ vs $k$ which are $k=3.07$ for the GRBs with $z_{obs}$ and $3.06$ for $z_{pred}$ showing that
both functions are consistent within 1$\sigma$ level.
The values of these $k$ are the ones for which the evolution is removed, and they correspond to the dashed blue line in the right panel of Fig \ref{luminosities} corresponding to $\tau=0$.
The additional dashed blue lines corresponding to $\tau= \pm 1$ denote the 1$\sigma$ values for the values of $k$.

\bibliographystyle{aasjournal}
\bibliography{refs}

\end{document}